\documentclass[sn-mathphys-num]{sn-jnl}

\usepackage{graphicx}%
\usepackage{multirow}%
\usepackage{amsmath,amssymb,amsfonts}%
\usepackage{amsthm}%
\usepackage{mathrsfs}%
\usepackage[title]{appendix}%
\usepackage{xcolor}%
\usepackage{textcomp}%
\usepackage{manyfoot}%
\usepackage{algorithm}%
\usepackage{algorithmicx}%
\usepackage{algpseudocode}%
\usepackage{listings}%
\usepackage[T1]{fontenc} 
\usepackage[utf8]{inputenc} 
\usepackage{color}
\usepackage{tabularx,booktabs}
\newcolumntype{Y}{>{\centering\arraybackslash}X}
\usepackage{caption}
\usepackage{subcaption}

\usepackage{url}
\usepackage{float} 
\usepackage{arydshln}

\begin{document}
\title[Analyzing and reducing the synthetic-to-real transfer gap in Music Information Retrieval: the task of automatic drum transcription]{Analyzing and reducing the synthetic-to-real transfer gap in Music Information Retrieval: the task of automatic drum transcription}

\author*[1]{\fnm{Mickaël} \sur{Zehren}}\email{mzehren@cs.umu.se}

\author[2]{\fnm{Marco} \sur{Alunno}}\email{malunno@eafit.edu.co}

\author[1]{\fnm{Paolo} \sur{Bientinesi}}\email{pauldj@cs.umu.se}

\affil*[1]{\orgdiv{Department of Computing science}, \orgname{Umeå university}, \orgaddress{\city{Umeå}, \country{Sweden}}}
\affil[2]{\orgdiv{Department of Music}, \orgname{Universidad EAFIT}, \orgaddress{\city{Medellín}, \country{Colombia}}}

\abstract{
      Automatic drum transcription is a critical tool in Music Information Retrieval for extracting and analyzing the rhythm of a music track, but it is limited by the size of the datasets available for training. A popular method used to increase the amount of data is by generating them synthetically from music scores rendered with virtual instruments. This method can produce a virtually infinite quantity of tracks, but empirical evidence shows that models trained on previously created synthetic datasets do not transfer well to real tracks. In this work, besides increasing the amount of data, we identify and evaluate three more strategies that practitioners can use to improve the realism of the generated data and, thus, narrow the synthetic-to-real transfer gap. To explore their efficacy, we used them to build a new synthetic dataset and then we measured how the performance of a model scales and, specifically, at what value it will stagnate when increasing the number of training tracks for different datasets. By doing this, we were able to prove that the aforementioned strategies contribute to make our dataset the one with the most realistic data distribution and the lowest synthetic-to-real transfer gap among the synthetic datasets we evaluated. We conclude by highlighting the limits of training with infinite data in drum transcription and we show how they can be overcome.
}



\maketitle
\bmhead{Acknowledgements}
The computations were enabled by resources provided by the National Academic Infrastructure for Supercomputing in Sweden (NAISS), partially funded by the Swedish Research Council through grant agreement no. 2022-06725.

\section{Introduction}

Music information retrieval (MIR) is a research field focused on extracting features from musical tracks. By retrieving a track's content, such as its structure, tempo, or key, we can build software to improve how we study or listen to music (e.g., providing content-based recommendations or facilitating music transcriptions for a variety of uses). In this context, several tasks in MIR are oriented toward the automation of musical, time-demanding operations that, until recently, we could execute only manually. One of them is automatic drum transcription (ADT). Specifically, ADT seeks to extract from an audio track the timing and instrument of the notes played on the drum kit. If this is already a difficult task because audio recordings are complex and difficult to analyze, it is even harder to do ADT in the presence of melodic instruments (DTM)~\cite{wu_review_2018} because the drum instruments to be transcribed are obfuscated by other audio sources.


Due to the difficulty of this challenge, the state-of-the-art methods often rely on deep learning (DL) models, as they are expressive enough to disentangle audio mixtures. However, because these models are usually trained in a supervised manner (i.e., with ground truth labels), their performance is hindered by the volume of the available datasets used for training. In fact, labeled datasets are scarce because audio tracks are labor-intensive to annotate, even for expert musicians, and the process is error-prone. Moreover, even in the case one makes the effort to create the annotations, tracks are often copyrighted and cannot be shared. Therefore, to counteract these issues, synthetic datasets have been generated.
In fact, by starting from a target musical score, such as a MIDI track, the associated audio can be reconstructed with synthesizers without manual labor and with perfectly accurate labels. 
Now, since MIDI tracks are abundant, it is possible to generate a virtually infinite amount of training data, or, at least, in quantities required by DL models. However, empirical evidence has shown that these synthetic datasets, despite their large size, are less effective than non-synthetic datasets in training models (e.g., \cite{zehren_high-quality_2023}). 

In fact, the lower standard, diversity, or complexity of synthetic datasets 
prevent the trained models from grasping all the nuances of real tracks, which is what the creators of the synthetic dataset SLAKH acknowledged: ``There is no getting around the fact that these are, in fact, MIDI files. [...] it's impossible to make the audio not sound cheesy if the MIDI files themselves are cheesy.''\footnote{\url{https://www.slakh.com/}} The issue of synthetic-to-real transfer has been extensively studied outside of ADT and two types of solution emerged to bridge the gap~\cite{nikolenko_synthetic_2019}: 1) Adapting the synthetic datasets so that they are more realistic or diverse (e.g., through system identification or domain randomization), and/or 2) adapting the learning algorithm to work on synthetic data (e.g., via transfer learning).
In this work, rather than adapting the learning algorithm to train on low-quality data, we study how synthetic datasets can be improved to reduce the transfer gap between them and real audio datasets.

Specifically, we highlight how the synthetic datasets for drum transcription are limited by their generation procedure and we propose a new dataset that does not suffer from the same pitfalls. 
In fact, three strategies make our generation procedure different from those used previously: 
1) Instead of following the more common practice of using MIDI files annotated offline, we increased the realism of the sequences of notes by employing human performances captured on electronic instruments.
2) Although a previous work~\cite{callender_improving_2020} already used human performances to synthesize drum-only audio, we did this with full tracks containing drums and accompaniment instruments, up to four of them.
3) Instead of the usual few dozens of synthesizer configurations (presets), we use hundreds of them to achieve a large diversity of timbres when rendering the audio.
With this generation procedure, we built a new synthetic dataset which is the only one to our knowledge to feature human performances, multiple voices and many presets, all in one dataset.

To evaluate the quality of our dataset and identify whether our generation procedure effectively narrows the gap between synthetic and real tracks more than what previous methods did, we conducted a set of three investigations.


First, we examined whether the different generation procedures can create realistic data distributions, that is, we compared the coverage of the real-world distribution with their synthetic distribution. This allowed us to discover where synthetic datasets lack in realism, as this lack represents the gap between synthetic and real, and may be the source of generalization errors.

Second, we quantified the transfer gap in terms of the training dataset. Specifically, by measuring how the real-world performance of the model scales by increasing the amount of training data (scaling laws), we could extrapolate the minimal loss we would expect to reach with each generation procedure.
This minimal loss will depend on the transfer gap between the training and test datasets.

Lastly, we performed an ablation study to validate the positive impact of the three main characteristics of our generation procedure on the transfer gap. To this end, as we did in the previous experiment, we compared different versions of our dataset: with and without human performances, with different numbers of accompaniment instruments, and different numbers of presets.

The remainder of this article is organized as follows. First, we present related works on the construction of datasets for DTM in Section~\ref{sec:related works}. 
Then, we introduce our solution to generate a better synthetic dataset followed by a comparison of the data distributions resulting from the different generation procedures, in Section~\ref{sec:training dataset}.
Finally, we describe our experimental design in Section~\ref{sec:experimental design} and we discuss the results of the comparison of the transfer gap between datasets and ablation study in Section~\ref{sec:results}.
We conclude in Section~\ref{sec:conclusions}.

\section{Related works}
\label{sec:related works}

In many fields of research relying on DL, the amount of training data is the main limiting factor.
This is also true for DTM where the datasets are usually very small, contain mistakes, or cannot be shared.

To solve this paucity of data, multiple approaches have been studied which we group into two categories: 1) semi-automatic annotations and 2) synthetic datasets.
The resulting datasets are compared in Table~\ref{tab:public datasets}.

\begin{table*}
      \caption{Scalar variables of seven DTM datasets. ADTOS is the dataset introduced in this manuscript.}
      \begin{center}
            \begin{tabularx}{\textwidth}{p{0.16\textwidth}YYYYp{0.0\textwidth}YYY}
                  \toprule
                                & \multicolumn{4}{c}{Synthetic audio} &                                 & \multicolumn{3}{c}{Real-world audio}                                                                                                                          \\
                  \cmidrule(r){2-5}
                  \cmidrule(l){7-9}
                  Dataset       & TMIDT \cite{vogl_towards_2018}      & SLAKH \cite{manilow2019cutting} & EGMD \cite{callender_improving_2020} & ADTOS        &  & RBMA \cite{vogl_drum_2017} & MDB \cite{southall_mdb_2017} & ADTOF-YT \cite{zehren_high-quality_2023} \\
                  \midrule
                  Tracks        & 4,228                               & 1,710                           & 1,059                                & 10,250       &  & 27                         & 23                           & 2915                                     \\
                  Duration (h)  & 260                                 & 115                             & $43\times10$                         & $2\times250$ &  & 1.6                        & 0.4                          & 250                                      \\
                  Voices        & n.s.                                & 4-48                            & 1                                    & 4            &  & -                          & -                            & -                                        \\
                  Drum presets  & 57                                  & 8                               & 43                                   & 512          &  & -                          & -                            & -                                        \\
                  Other presets & 1                                   & 179                             & 0                                    & 458          &  & -                          & -                            & -                                        \\
                  \bottomrule
            \end{tabularx}
      \end{center}
      \label{tab:public datasets}
\end{table*}

\subsection{Semi-automatic annotations}
While some datasets have been fully annotated by hand, such as RBMA~\cite{vogl_drum_2017}, the general approach is to employ semi-automatic techniques to reduce the manual labor required to create the labels.

One common technique is to bootstrap the annotations to an intermediate state
and then use human annotators only to refine the labels to the desired level of detail or precision.
For example, both ENST and MDB were created with onset detection algorithms applied to audio files of isolated instruments (stems). Then, the annotators verified and labeled the initial set of positions to the correct instruments~\cite{gillet_enst-drums:_2006,southall_mdb_2017}.
Similarly, in the related field of multi-instrument transcription, Simon et al. used monophonic recordings, for which labels can be precisely computed with a pitch detection algorithm, and mixed them into polyphonic tracks~\cite{simon_scaling_2022}. Since the merged audios come from unrelated recordings, the data is not musically coherent. However, thanks to its large diversity, this set made the model generalize better when pre-trained on it.

Other datasets were built with the opposite approach: refining human annotations with the help of algorithms. This technique is notably required with crowdsourced datasets where annotations can vary in quality. It is commonly used to align precisely the annotations to the audio recordings, a task that can be difficult or tedious for humans when the recording varies in tempo, as well as to estimate a quality score of the annotations and, thus, filter out problematic tracks.
A2MD was created in this manner, by searching and aligning MIDI files found in the wild to their corresponding audio recording on Youtube~\cite{wei_improving_2021}. The alignment was performed by warping each MIDI file to minimize its difference, when synthesized, with respect to the original audio. Then, based on the quality of the alignment, the annotations were grouped into three levels of fidelity. 
It is interesting to note that a dataset of piano recordings such as MusicNetEM was created with a similar approach to A2MD, with the difference that the annotations were aligned to an initial algorithmic transcription, rather than to the audio~\cite{pmlr-v162-maman22a}. 
After comparing the two kinds of alignments, the authors claimed that the alignment in the symbolic domain is more robust as it can ignore errors in the original transcription. 
Lastly, ADTOF-RGW and ADTOF-YT repurposed annotations meant for rhythm video games to music transcription by refining the alignment of the annotated beats to audio cues detected by a beat tracking algorithm \cite{zehren_adtof_2021,zehren_high-quality_2023}. 
This approach proved to work well, but requires a rough initial alignment~\cite{driedger_towards_2019}, after which the labels are automatically converted from the video game instruments to their real-world counterpart with a rule-based system. Finally, to further clean the annotations from human mistakes, a manual check was performed on those tracks where a preliminary algorithmic transcription achieved a low score.


\subsection{Synthetic datasets}
Rather than making the annotation process easier, 
synthetic datasets aim at scaling the amount of data by automating its generation. Once the generation process is in place it has the benefit of producing large amounts of data at a very low cost.
Moreover, the generation process can be tuned to control the data distribution, which can either be used to increase diversity (e.g., to complement existing real-world datasets by covering underrepresented occurrences), or to perform controlled experiments (e.g., to perform an ablation study). Lastly, the generation of original tracks is not affected by copyright constraints. Despite such advantages, the question of domain adaptation arises, that is, one could wonder how well a model trained on a synthetic data distribution will perform in the real world.

The issue of domain adaptation can be tackled in two ways.
First, by improving the data distribution of the synthetically generated dataset to be close to that of a dataset of real tracks, which is something very difficult to achieve given the level of complexity of real audio tracks (also, it is not known to what extent realism, as opposed to diversity from the randomization of the generation procedure, is necessary~\cite{nikolenko_synthetic_2019}).
Second, by adapting the algorithm to effectively learn from synthetic datasets and transfer to the real domain.
As an illustration of these two techniques, we now present different synthetic datasets.



TMIDT, published in 2018, was generated from 4194 MIDI files (259h) of pop tracks from a freely available online collection\cite{vogl_towards_2018}. 
To ensure diversity of sounds, each file was synthesized with one of 57 SoundFonts (the equivalent of synthesizer presets) for the drum kit, but only one general synthesizer for the non-drum instruments. 
Moreover, the dataset is available in a normal version and a balanced version where the MIDI files have been altered to balance the occurrences of the drum instruments, effectively covering instruments that are not sufficiently represented in the normal version. Although this modification has been done musically (i.e., with coherent permutations of drums and cymbals), the artificially generated patterns did not help the generalization of the model. Therefore, balancing the dataset did not have a positive effect when testing on real tracks.
To help the model transfer to real-world audio, multiple training scenarios have been evaluated. Overall, models trained exclusively on synthetic data can generalize to some extent to real-world datasets. However, better performances have been achieved with transfer learning, i.e., by having the model pre-trained on synthetic data and then refined on real-world data. 

SLAKH (redux), published in 2019, was generated from 1709 MIDI files (115h) from the Lakh MIDI dataset~\cite{manilow2019cutting,raffel_learning-based_2016}. Compared to TMIDT, fewer (8) synthesizer presets have been used to render the drum, and more (179) have been used for the accompaniment. 
The MIDI files contain a maximum of 48 voices and a minimum of 4: piano, bass, guitar, and drums. 
By controlling which voice to include in the mix~\cite{manilow_simultaneous_2020}, Manilow et al. evaluated the impact that the number of instruments playing has on the transcription and separation tasks. They found that while the recording becomes more complex, with an increasing number of voices, the transcription also gets more difficult to achieve for all the instruments but the drums and the bass.

EGMD, published in 2020, has been synthesized from 1058 human performances (10h) captured on an electronic drum kit~\cite{callender_improving_2020,gillick_learning_2019}. Thus, compared to the other datasets which are annotated offline, the timing and dynamics of the notes are expected to be more human-like. However, no accompanying instruments are included and only 43 drum synthesizer presets have been used to render each score (for a total of $43 \times 10$h). This resulted in identical sequences redundantly shared by different tracks, a situation that required data augmentation to successfully train the models.

Lastly, we mention AAM, published in 2023, which consists of 3000 musical scores (125h), generated with an automatic procedure based on music theory and synthesized in a similar way to SLAKH~\cite{ostermann_aam_2023}. However, because the aim in the creation of this dataset was on the fast generation of a large number of tracks, the authors acknowledge that the drum patterns are very simplistic. Thus, we did not include it in Table~\ref{tab:public datasets}.

\section{Training dataset}
\label{sec:training dataset}

The datasets TMIDT, SLAKH, and EGMD have been synthesized with different procedures and have different limitations. To study the impact of these limitations on the performance of the trained models and remove them, we followed a new procedure to generate a synthetic dataset that we named Automatic Drum Transcription On Synthesizers (ADTOS) and that has the following characteristics all at once: a large number of MIDI tracks, human performances with accompaniment instruments, and many synthesizer presets. This dataset is also presented in Table.~\ref{tab:public datasets}. Next, we describe the generation procedure and then we compare the datasets in terms of their data distributions.

\subsection{Generation procedure}

Similarly to the other synthetic datasets, ADTOS' tracks are built from a symbolic representation of notes in MIDI, which is then rendered into audio by synthesizers. Our procedure works as follows: First, we collected MIDI loops of performances captured on electronic instruments; then, we assembled the loops to compose tracks with drums and accompanying instruments; finally, we synthesized the tracks with virtual instruments and a wide variety of presets. We now illustrate these three steps in more detail.

To gather the symbolic representations of notes, we used professional MIDI loops that were expressly generated as building blocks to compose tracks.
These loops are particularly suited for a dataset as they consist of many short sequences of drums (about 130,000 loops, 250h), piano (about 20,000 loops, 100h), and bass guitar (about 7,000 loops, 30h), are grouped by themes to build cohesive tracks and cover a wide variety of sections, genres, time signatures, and tempi. 
Further, the loops are recorded by professional session players on electronic instruments. Thus, as opposed to an offline creation process where the notes are fully quantized, these performances contain slight variations in the notes' position and dynamic. 

To build full tracks from these MIDI loops, we assembled the MIDI snippets iteratively, that is, we built a track section by section, in sequence, and a section is built by layering loops for each voice. This process is repeated until reaching the desired duration and number of voices.
Specifically, to ensure some degree of coherence in the composition, a track's section is initialized with a ``master'' MIDI loop selected from the same theme as the previous section. Then, the section is built by layering loops with similar tempo, length, genre, time signature, and key signature on top of the master. 
This process ensures that the different sections of a track are variations of the same theme, possibly with different characteristics (e.g., a tempo change), while the loops within a section have uniform characteristics (e.g., the same tempo as the master).
Loops are selected with replacement (a loop can be used more than once) based on a distance function that takes into account both the compatibility of the loop with the desired characteristics and how many times it has already been selected. 
This effectively allows us to generate any number of tracks by evenly sampling the MIDI loops available and creating different combinations of overlaps when files are reused. With this method, we generated 10,250 MIDI tracks \footnote{We chose this number of tracks to build a training set with $2^{13}$ tracks}. Each track features drums, piano, bass guitar, and guitar loops \footnote{Because we do not own loops for the guitar, we simulated this instrument with loops meant for the piano.}. In particular, drum loops are used twice on average for a total duration of $2\times250$h.

Finally, to transform the tracks into audio, we followed the approach used to curate SLAKH. 
We rendered each instrument with its corresponding synthesizer and a random preset.\footnote{This step is by far the most time-consuming of the whole generation procedure. Synthesizing the audio for a single instrument is about ten times faster than real-time when running in parallel on a MacBook Pro 2019.}
Because the number of tracks we composed was larger than the number of presets, each preset was used multiple times. However, we made sure to reuse it with different presets for the other instruments, thus increasing diversity in timbres.
Finally, before merging all the tracks into a single audio file, we normalized the loudness of files of isolated instruments with the EBU R128 algorithm available in Essentia~\cite{bogdanov_essentia_2013,noauthor_loudness_2011}. A total of 512 drum presets and 458 non-drum presets were used. 

\subsection{Comparing data distributions}
By generating audio as close as possible to real tracks, and thus reducing the synthetic-to-real gap, we were able to curate the largest and most diverse dataset currently available.
However, having a large quantity of information does not necessarily translate into a better training of the model if this information is not meaningful.

\begin{figure*}
      \centering
      \includegraphics[width=\textwidth]{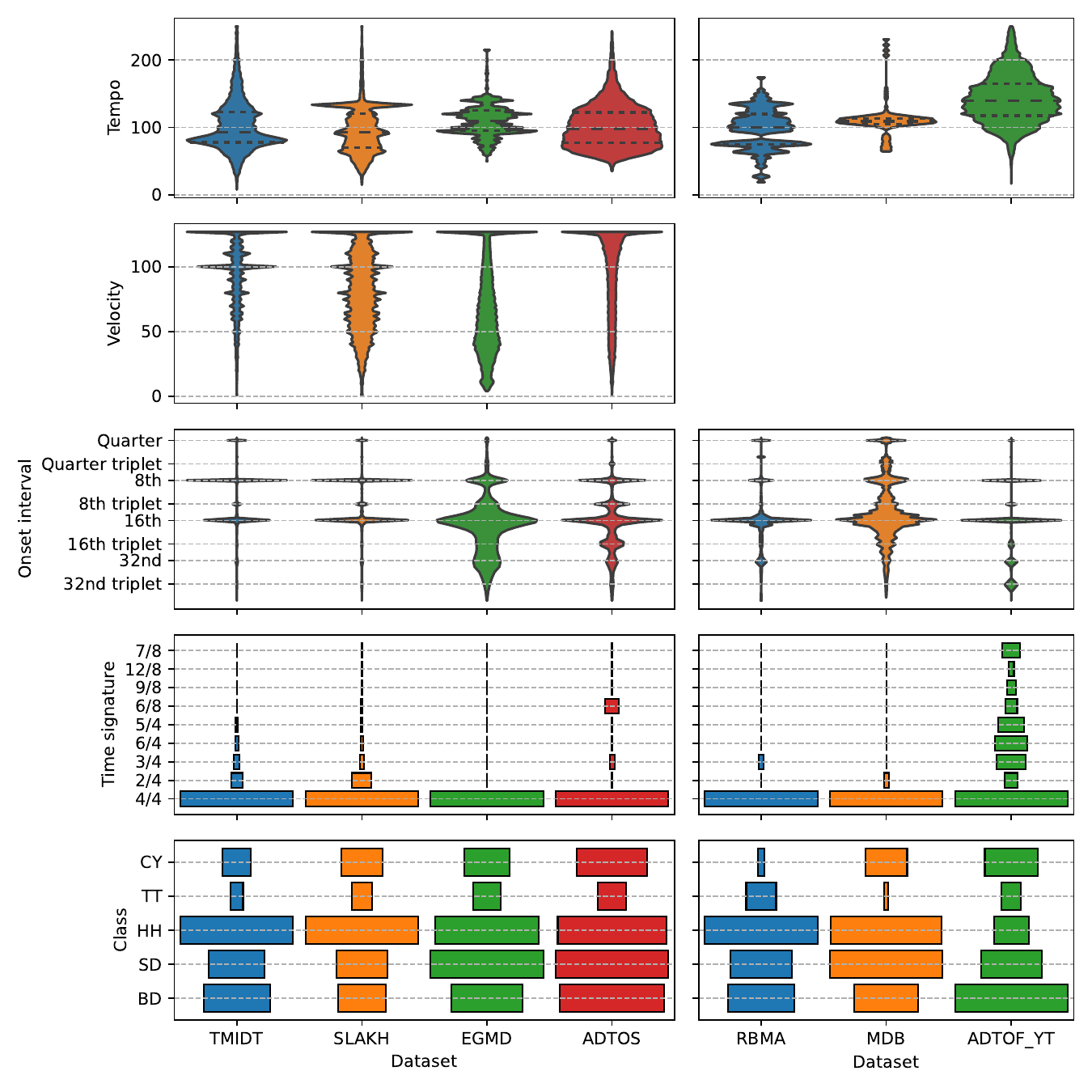}
      \caption{Violin and bar plots representing respectively continuous variables distributions (tempo, velocity, and onset interval) and discrete variables distributions (time signature and class), for the synthetic datasets (left column) and real-world datasets (right column). The distributions are normalized, so that each plot has the same width.}
      \label{fig:distributions}
\end{figure*}

Consequently, to assess the quality of ADTOS, we compared it with the other datasets.
More specifically, we overlapped and compared their distribution, because different generation procedures result in different data distributions and because a source distribution can only train a model that performs well on a similar (or subsumed) target distribution.
Since our synthetic generation procedure is a cheaper alternative to datasets curated from real-world tracks, we were also interested in comparing synthetic to real-world distributions in order to highlight differences that are possible causes of generalization errors. Both kinds of distributions are respectively plotted on the left and right side of Figure~\ref{fig:distributions}.
We now describe and compare the distributions of the five variables pictured in the rows of Figure~\ref{fig:distributions}: tempo, velocity, onset interval, time signature, and class.

\paragraph*{Tempo,} measured in beats per minute (bpm), dictates the speed at which music is played, with higher tempi indicating smaller note intervals (for the same beat subdivision). We observe two key differences in the distributions of tempi among the datasets.
First, the lack of representation of some tempi, especially the lack of fast tracks in all the datasets but ADTOF-YT, suggests a potential source of generalization error.
Second, the tempo of the tracks follows a normal distribution in the large real-world dataset ADTOF-YT (RBMA and MDB are too small to confirm this trend for all real-world datasets), but not in most synthetic datasets. 
Indeed, only ADTOS is close to a normal distribution while TMIDT, SLAKH, and EGMD have peaks around some values (e.g., 90 bpm or 120 bpm). 
Although it is not clear if those peaks are there to represent meaningful tempi in specific music genres or are simply due to limitations in the generation procedure, it is anyway an indication that increasing the dispersion of tempi in those datasets might improve generalization.

\paragraph*{Velocity} indicates how hard the instrument is struck to play a note (dynamic), and ranges from 0 to 127 in MIDI, with 127 being the loudest.
Unfortunately, there are no accurate velocity annotations on real-world datasets. Still, we can compare synthetic distributions to identify trends.
In fact, all the distributions are skewed toward the maximal velocity and have a long tail towards lower values. Despite such a similarity, the long tail is not the same in every dataset:
datasets annotated offline (TMIDT and SLAKH) have peaks around specific values (e.g., 100), whereas datasets captured on electronic drum kits (EGMD and ADTOS) are more dispersed. We assume that the peaks are due to the quantization of dynamic levels in the software used to create the annotations (e.g., "forte" might correspond to a MIDI velocity of 100).

\paragraph*{Onset interval}
represents the distance, relative to the beat interval, between two consecutive drum onsets.\footnote{To compute this distance, we merged all notes taking place in a 50ms window, as we assumed they were meant to be played in unison, and we discarded the values above quarter notes.}
We can see two trends in the datasets.
First, although there are clear peaks around common beat subdivisions, some of them are not present in all the datasets. Especially faster subdivisions (e.g., 16-note triplets or 32nd notes) are missing in TMIDT and SLAKH, which is possibly due to simplistic MIDI files.
Second, a spread around the common subdivisions is present in captured annotations, both from real-world distributions (MDB, annotated with an onset detection algorithm) and synthetic distributions (EGMD and ADTOS, recorded on electronic drum kits), but not when the annotations are created offline, because in this way an annotation quantizes the note intervals to the exact beat subdivisions (as it happens with velocity), whereas the capture will preserve human deviations. This is a second indication that TMIDT and SLAKH MIDI might be too simplistic.

\paragraph*{Time signature} establishes how the beats of a track are organized in bars and how notes in a bar are grouped. In other words, it tells how rhythm is eventually perceived.
While many time signatures exist, 4/4 is by far the most common in Western commercial music (e.g., Rock, Metal, EDM), which is the focus of these datasets. Thus, it is not surprising that 4/4 has the most occurrences in all datasets. However, it is important to note that ADTOF-YT has a large diversity of time signatures, most of which are not covered by the other datasets. This is an indication that the other datasets are more biased than ADTOF-YT.

\paragraph*{Class} distribution tells what (group of) instruments are more commonly occurring in each dataset. 
Since datasets have a different number of classes, to compare them we mapped their vocabulary down to the largest common vocabulary, which is the 5-class vocabulary from ADTOF-YT. This effectively grouped some instruments to the same class (e.g., a ``crash cymbal'' or a ``ride cymbal'' are both mapped to ``CY''), as detailed in our previous work~\cite[p.821]{zehren_adtof_2021}. 
While some of those classes are sparse in real-world datasets (i.e., ``CY'' in RBMA and ``TT'' in MDB), the fact that they all appear with high frequency in the synthetic dataset should facilitate generalization.

\section{Experimental design}
\label{sec:experimental design}

Overall, ADTOS is closer to the real-world distribution compared to the other synthetic datasets (we further investigate this trend in terms of drum sequences in Appendix~\ref{appendix A}), which is a good sign that it might have a smaller synthetic-to-real gap. However, this conclusion is based on the evaluation of only five features extracted from the drum voice of the MIDI files, which are not representative of all the characteristics of the audio, for example, they do not take into account the synthesizer presets or the other instruments. Furthermore, we do not know to what extent our better coverage effectively correlates to improvements in the model's performance.

Therefore, to quantify the transfer gap, we estimated the minimal generalization error we can achieve when training on data generated with each procedure. 
However, since the trained model might benefit from an unknown amount of data points, and because it is always possible to generate more synthetic data, we cannot measure the minimal generalization error with respect to a fixed data size.
Instead, we can apply a scaling law, that is, we can study the scaling of the models' performance when increasing the quantity of training data.

Scaling laws are a well-studied phenomenon used to link different properties of an algorithm to the performance of a model (a survey is available in Villalobos~\cite{epoch2023scalinglawsliteraturereview}).
The link is used to predict a dependent variable, usually the performance of the model measured in cross-entropy loss, when tuning independent variables such as the number of training samples, the number of parameters in the model, or a computation budget.
As an example, by estimating the parameters of the scaling law from empirical evidence, it is possible to predict how large the training dataset needs to be to reach a desired accuracy~\cite{hestness_beyond_2019}. 
After adjusting the number of parameters and the computation budget so that they are not limiting factors, the loss of the model $L$ achieved by training on a quantity of data $n$ will form a learning curve that follows a scaling law~\cite{hestness_deep_2017}:
\begin{equation} 
      \label{eq:scaling}
      L(n)\simeq \alpha n ^{-\beta} + \gamma
\end{equation}
Where $\alpha > 0$ is a constant property of the problem, $\beta > 0$ is the scaling exponent that defines the steepness of the learning curve, and $\gamma \geq 0$ is the lower bound of the loss.
Concretely, the learning curve shows how fast a model improves from adding training data, and at what value it will plateau.\footnote{The learning curve represents the loss of a model trained until convergence in terms of the training data size. Thus, it is different from training and validation curves (see~\cite{hestness_deep_2017}).}

This lower bound, $\gamma$, depends on two key characteristics of the algorithm: the irreducible error and the transfer gap. On the one hand, the loss of the model is bounded by the irreducible errors specific to the test data, i.e., Bayes errors or errors due to incorrect labels in the ground truth\cite{hestness_deep_2017}. On the other hand, the loss of the model is also bounded by the gap between the training and test datasets~\cite{amini_scaling_2023}.
With that in mind, the idea behind our experiments was to change the training data while keeping the test data constant. This way, we tempered the impact of the irreducible errors so that the difference in $\gamma$ is only explained by the transfer gap.
We now describe in more detail the independent, dependent, and control variables we used to study the scaling laws.

\subsection{Independent variable}
To vary the quantity of training data, we trained a model by increasing the number of tracks from 1 to 8192. 
Although the number of tracks is different from the de facto variable of the number of samples used to study scaling laws ---tracks have different durations and, thus, contain different numbers of samples--- we believe that this gives a better notion of the quantity of information in the training data. In fact, since a track is uniform in timbre and has repeated sections, increasing the number of samples while keeping the same number of tracks will only add redundant information.


In order to evaluate scaling laws of synthetic audio, we trained a model exclusively on synthetic data, as Fan et al.~\cite{fan_scaling_2023} did. Note that this practice deviates from the common case where synthetic datasets are employed in addition to real datasets, for example to cover parts of the data distribution that are not sufficiently represented (e.g.~\cite{amini_scaling_2023}). In fact, the performance of the trained model may not depend on the quality of the synthetic dataset, but on how well that dataset complements real ones.
Instead, by training only on synthetic data we can estimate a more general effectiveness of the generation procedure.
We assume that any improvement in the generation procedure will also impact models trained with a mix of synthetic and real data.


\subsection{Dependent variables}
To estimate the performance of the model at each training size, we had to decide 1) the metric used to quantify the performance of the model and 2) the datasets to use for testing.

The performance of a drum transcription is usually computed with the F-measure, which is the harmonic mean of precision and recall of the estimated drum onsets. An estimation is considered correct if it is within a small window (e.g., 50ms) from the ground truth. 
Although intuitive, this score is imprecise and not a direct measure of the performance of the model (e.g., the F-measure does not capture fine information, like noise in the activation, in the output of the model) as it depends on the quality of the peak-picking procedure for the discretization of the model's activation into onsets.
Therefore, instead of the F-measure, we used the difference between the model's activation and the ground truth (the loss), computed with the widely used and more precise binary cross-entropy metric. Although the loss does not have an interpretation in absolute terms, a relative reduction in loss means that the model is improving.

To measure the models' loss, we use both validation and test datasets. 
The validation loss is measured on tracks generated from the same procedure as the training data (with unseen MIDI and presets) and indicates the ``on-domain'' performance of the model. Although the on-domain performance does not express the synthetic-to-real gap, it is anyway interesting to understand the dynamic of the system. Particularly, this helps us quantify how much the improvements in the validation loss transfer to the test loss~\cite{abnar_exploring_2022}. Moreover, the validation loss can tell us about the complexity of the training dataset~\cite{amini_scaling_2023}.
The test loss, instead, is measured on a mixture of datasets containing exclusive real-world audio and indicates the generalization capability of the model in the real domain. To cover a wide range of music, we tested on four datasets: ADTOF-RGW~\cite{zehren_high-quality_2023}, ENST~\cite{gillet_enst-drums:_2006}, MDB~\cite{southall_mdb_2017}, and RBMA~\cite{vogl_drum_2017}.

\subsection{Control variables}
As the training data grows, different variables have to be controlled, so that they will not influence the outcome of the experiments. We will talk here about 
the size of the model, the number of training steps, and the training strategy, as these variables were those that had the strongest impact on the results of the experiments.

The number of parameters of the model has to grow in size to accommodate for the added information when the training data increases (e.g.~\cite{droppo_scaling_2021}), otherwise the model will saturate and will not be able to learn. However, it is not known apriori at what ratio the number of parameters has to increase with respect to the data size, as the ratio depends on the task and the increase of parameters might not be linearly correlated with the increase in the number of tracks. Thus, to control the size of the model we tested a range of values (between $2\times10^5$ and $1.6\times10^7$ parameters) and we kept the score of the best-performing architecture at each training size. This ensured that the number of parameters did not represent a limit throughout the experiments.

Likewise, we tuned the computation budget so that it would not be a limiting factor. Because different models and datasets require different numbers of training steps, we used ``early stopping'' to train models until convergence. Although it has been shown that training a model to convergence is inefficient (e.g.~\cite{droppo_scaling_2021}), this warrants to find the smallest loss possible for any combination of parameters and training sizes.

Lastly, we fixed different hyperparameters to secure proper convergence of the models throughout the experiments.
We found that evaluating the validation loss for early stopping at each complete pass of the training dataset (epoch) did not work, since we trained on data varying over multiple orders of magnitude in size and early stopping did not have enough patience to guarantee the convergence of the model in the low data regime. To solve this issue we fixed the validation interval to 128 training steps instead of an epoch and we stopped the training only after no improvement had been noticed for 25 validations in a row.
The steps are composed of a batch of $32$ audio snippets. The audio snippets consist of 4s of audio, taken from random tracks, which are transformed into a mel-scale spectrogram with a resolution of 100Hz and 12 bands per octave between 20Hz and 20kHz.
Finally, we used Adam optimizer and a learning rate scheduler with a warm-up phase, with a learning rate increasing from $0$ to $5\times10^{-4}$ for $128$ steps and then decaying by a factor $\times10$ over $8192$ steps.



\section{Results}
\label{sec:results}

We used the design presented in the previous section to realize two experiments.
First, we quantified the transfer gap on three datasets to establish a reference for different generation procedures. Then we conducted an ablation study to evaluate the impact of the different characteristics of our generation procedure on the transfer gap.

\subsection{Transfer gap for different generation procedures}

\begin{figure}
      \centering
      \includegraphics[width=0.66\columnwidth]{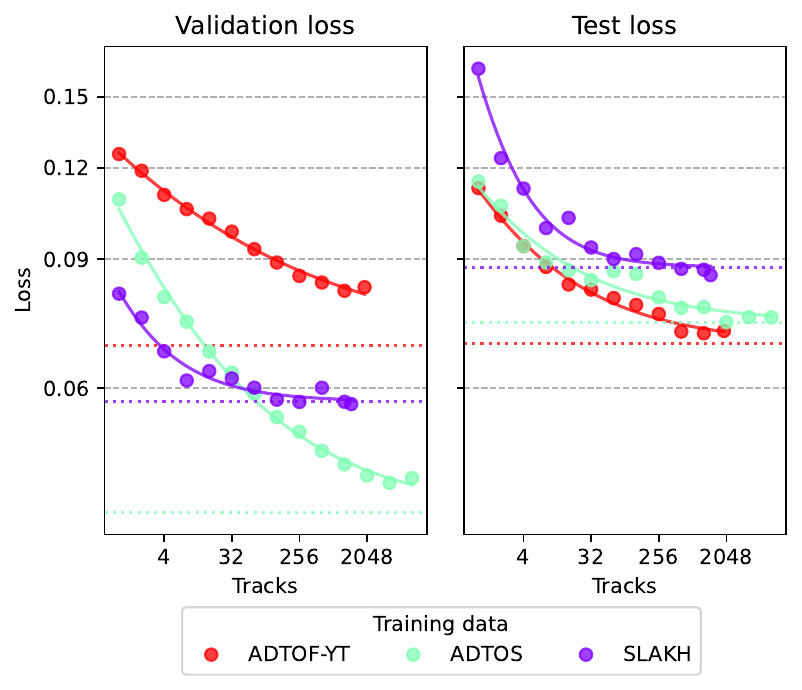}
      \caption{Validation and test loss in function of the number of tracks when training on different datasets. The solid lines represent the learning curves, fitted in the log space, from equation~\ref{eq:scaling}. The dashed lines represent the value of $\gamma$, the lower bound of the loss. Notice the log-log scale.}
      \label{fig:reference scaling}
\end{figure}

In Figure~\ref{fig:reference scaling}, we compared ADTOS with SLAKH and ADTOF-YT, which served as baselines for the synthetic and the real-world data, respectively.
Note that since ADTOF-YT contains data from the real domain, training on this dataset is not representative of the synthetic-to-real transfer.
In fact, testing on a new distribution of the same domain is considered easier than testing from the synthetic to the real domain. 
Thus, we trained on ADTOF-YT to estimate an ideal value of the transfer gap, that is, the smallest error we hoped to achieve with synthetic data.

Training on ADTOF-YT leads to the smallest lower bound of the test loss (red dashed line in the right plot of Figure~\ref{fig:reference scaling}), thus confirming that this dataset has the smallest transfer gap.
Instead, training on SLAKH leads to the highest transfer gap (purple dashed line in the right plot), while ADTOS stays in between (green dashed line in the right plot). 
This indicates that our generation procedure yields a dataset that performs better than the other synthetic dataset, although not as good as a dataset of real audio tracks.

Although training on ADTOF-YT gives the smallest lower bound of the test loss, it also leads to the highest lower bound of the validation loss (red dashed line in the left plot).
This means that ADTOF-YT has the highest irreducible errors which, we recall, are due to either errors in the labels or Bayes errors. This is not surprising considering that the synthetic datasets (green and purple dashed lines in the left plot) are generated artificially and, therefore, do not have errors in the labels.  
Moreover, since the synthetic datasets are likely less complex than real audio (e.g., there are no audio effects or vocals), fewer elements obfuscate the onsets in the signal, thereby reducing the amount of Bayes errors.

We also noticed that the learning curves from training on ADTOF-YT and ADTOS did not reach the lower bound for the validation loss (red and green solid lines in the left plot). 
This means that, in these two cases, the models are far from the irreducible error and will gain from more training data.  
However, this is not the case for SLAKH (purple solid lines in the left plot), most likely because in this dataset the presence of redundant tracks causes a lack of diversity that ultimately makes the learning curve plateau after a few tracks. In the test loss, instead, all the learning curves reached a plateau, which is an indication that
further training data may reduce the validation loss, but will not transfer to a better performance on the test datasets.

In summary, we deduced that the datasets are not limited by their size, but only by the diversity they achieve from their generation procedure. Although our generation procedure is not as good as training on real data, it is better than the other synthetic procedure we evaluated.


\subsection{Ablation study}
\begin{figure}
      \centering
      \begin{subfigure}{0.66\columnwidth}
            \includegraphics[width=\columnwidth]{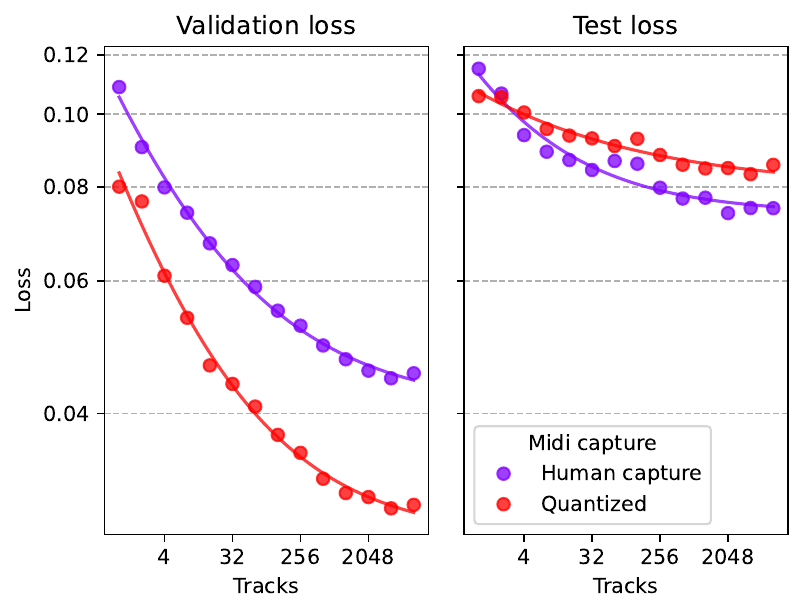}
            \caption{MIDI source}
            \label{fig:ablation study midi source}
      \end{subfigure}
      \begin{subfigure}{0.66\columnwidth}
            \includegraphics[width=\columnwidth]{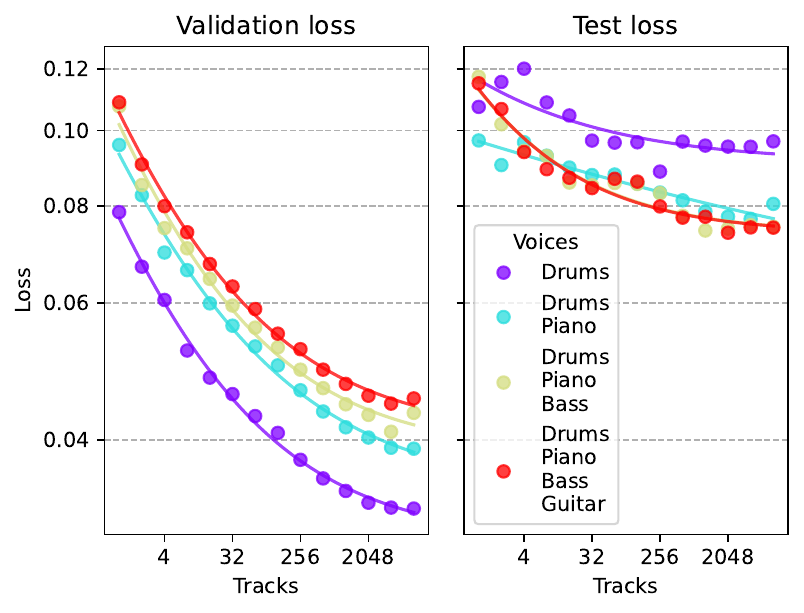}
            \caption{Voices}
            \label{fig:ablation study voices}
      \end{subfigure}
      \begin{subfigure}{0.66\columnwidth}
            \includegraphics[width=\columnwidth]{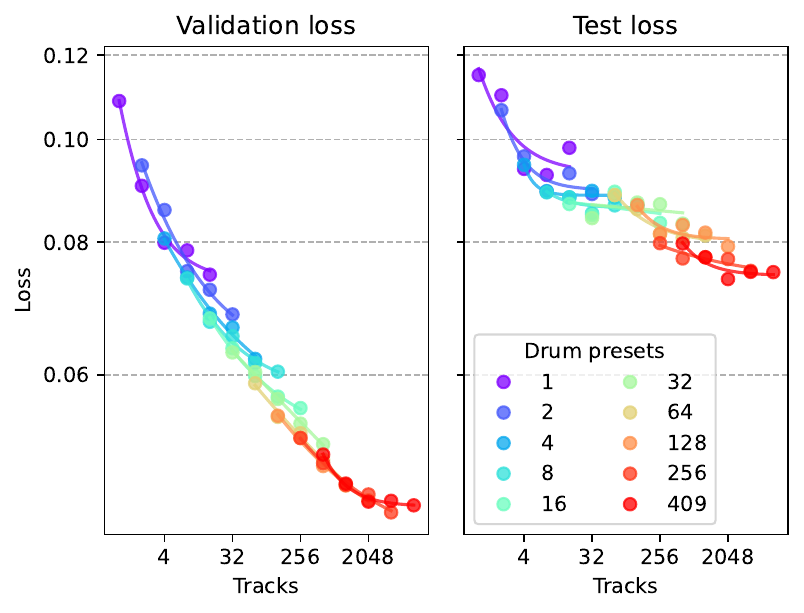}
            \caption{Drum presets}
            \label{fig:ablation study presets}
      \end{subfigure}
      \caption{Learning curves for different versions of ADTOS by modifying: a) the MIDI source, b) the number of voices, and c) the number of presets.}
      \label{fig:ablation study}
\end{figure}

To validate that the characteristics of our generation procedure are effectively reducing the transfer gap, we conducted an ablation study by evaluating different versions of our dataset, as we did in the previous experiment. 
We investigated the three main characteristics of our generation procedure: the MIDI captured from human performances, the accompaniment instruments, and the number of drum presets. They are presented in Figure~\ref{fig:ablation study}.

\paragraph*{Human performance} In order to compare in a fair way the training on MIDI files of human performances with the training on MIDI annotated offline, it is necessary to keep everything else constant (e.g., same tracks, same presets). 
However, since we had only the human performance version of the tracks at our disposal, the sole way of keeping everything else constant was to simulate the offline annotation process with MIDI captured from the drum instruments.
To simulate the offline annotations, we quantized the timing and velocity of the performances according to the distributions of the offline annotations from TMIDT and SLAKH presented in Figure~\ref{fig:distributions}. 
Specifically, we quantized both the notes to the grid of 16th notes, the smallest beat subdivision in the dataset, and the velocities to the values 127 and 100, the two most common values. The comparison between the original and quantized version of our dataset is presented in Figure~\ref{fig:ablation study midi source}.

While the quantized version has a lower validation loss compared to the original version (red line compared to purple line in the left plot), it also has a higher test loss (red line compared to purple line in the right plot).
This indicates that, compared to the simulated offline annotation, capturing human performance to create MIDI increases data complexity and reduces the transfer gap. Hence, the validation loss increases and the test loss decreases.

\paragraph*{Accompaniment instruments} To evaluate the effects of the accompaniment instruments on the transfer gap, we used the same MIDI tracks to generate four versions of the datasets with an increasing number of voices. 
As represented in Figure~\ref{fig:ablation study voices}, we evaluated the training of the models with one, two, three and four voices by iteratively adding piano, bass guitar, and guitar to the drums.

While the validation loss increases with more voices (left plot, going from purple to red), the test loss decreases (right plot, going from purple to red).
Similarly to what happened in the previous evaluation, the more complex the training data become by adding new instruments to the audio the lower the transfer gap. 
However, while we note a large difference between recordings containing only drums and drums with piano, there are diminishing returns when voices are more than two (the curves get closer to each other), to the point that there is no difference in test loss between three and four voices (the green and red curves overlap in the right plot). Considering that real-world audio tracks usually contain many more instruments, this is an indication that our tracks are too simplistic in the way they overlap voices.
Nonetheless, our semi-coherent overlap of instruments is still better than drum-only recordings.

\paragraph*{Number of drum presets} Lastly, to evaluate the benefit of having a large number of presets, we trained the models on truncated versions of our datasets, as presented in Figure~\ref{fig:ablation study presets}.
Specifically, since the dataset is synthesized by reusing presets for different MIDI files, we trained the models by either increasing the number of presets or the number of times each preset is used.
Thus, each fitted curve represents the scaling achieved, from a fixed number of presets, by increasing the number of times each preset is used. 
Note that since each preset is used between 1 and 20 times, the number of tracks in the dataset depends on the number of presets. 

We noticed that the validation and test losses attained with any number of presets first decrease and then plateau above the loss of the next number of presets. In fact, all the curves are going flat above the following curve.
This means that, regardless of the total number of tracks generated, reusing the presets more than two to four times has diminishing returns.
Likewise, increasing the number of presets for the same number of tracks (i.e., reducing the number of times each preset is reused) decreases both losses. 
We conclude that increasing the number of presets effectively increases the diversity of the dataset, which helps reduce the transfer gap.

In summary, the three characteristics of our dataset, by either increasing the data complexity or its diversity, effectively narrow the synthetic-to-real gap.

\section{Conclusions}
\label{sec:conclusions}

While the use of synthetic datasets in DTM has the potential to solve issues related to small dataset sizes, annotators' mistakes, and copyrights, these datasets have been inefficient at training models able to transfer to real audio tracks.
In fact, by analyzing existing generation procedures, we identified that synthetic data are possibly limited by their simplistic MIDI files, the number of presets used, or the lack of non-drum instruments in the MIDI containing human performances. 
To overcome those limits, we proposed to use professional-grade MIDI loops of human performances captured on drums, piano and bass guitar and more presets than previously experimented to build complex and diverse tracks with up to four voices. Then, we showed that this generation procedure has, indeed, a more realistic data distribution on multiple variables.
Further, we confirmed that the transfer gap achieved by our dataset is better than the other synthetic dataset we evaluated, although not as good as training on real-world audio.
Finally, we validated with an ablation study that all the advantages of our generation procedure, compared to the previous ones, are effectively reducing the transfer gap.

With this work, we proved that it is not the amount of data that matters for datasets in DTM, but its quality, which depends on the generation procedure. Overall, the more complex and diverse the training data, the smaller the transfer gap. 
To pursue this direction, in future works we would like to assess a finer scaling of the performance of the model (e.g., on different drum instruments or music genres) to identify points of improvement in the generation procedure.

\backmatter

\begin{appendices}
      \section{Drum sequences coverage between datasets}
      \label{appendix A}
      
      \begin{figure*}
            \centering
            \begin{subfigure}[b]{0.49\textwidth}
                  \centering
                  \includegraphics[width=\textwidth]{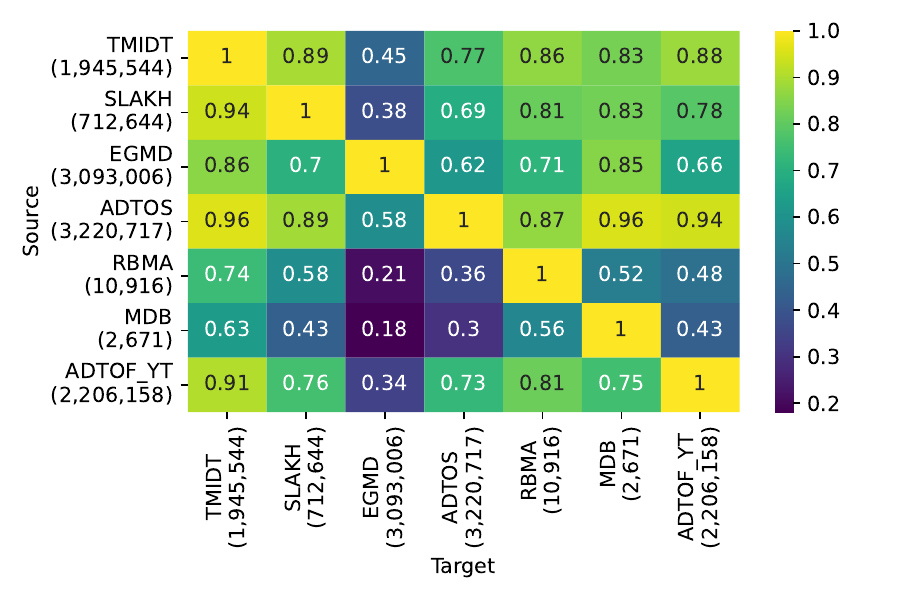}
                  \caption{Coverage in beats}
                  \label{fig:BeatsOverlap}
            \end{subfigure}
            \hfill
            \begin{subfigure}[b]{0.49\textwidth}
                  \centering
                  \includegraphics[width=\textwidth]{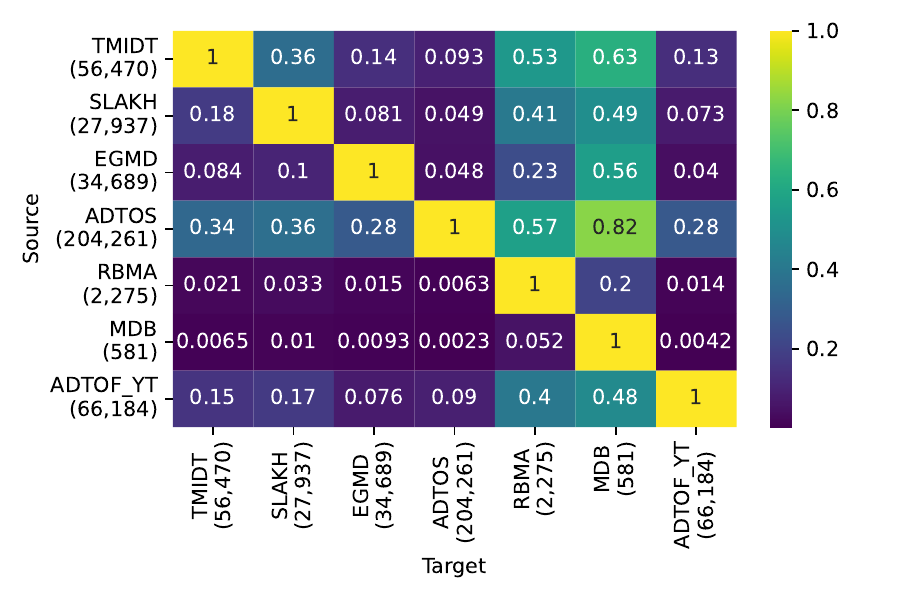}
                  \caption{Coverage in unique beat sequences}
                  \label{fig:GroupsOverlap}
            \end{subfigure}
            \caption{Relative frequency at which beats (left) or unique beat sequences (right) from a target are included in the source. Numbers in parentheses represent the count of beats or sequences in the datasets.}
            \label{fig:overlap}
            
      \end{figure*}
      
      Due to the repetitive nature of music, the duration of a dataset is not a precise indication of the quantity of information it contains. Instead, we argue that the variety of tracks is a better indication since different tracks have different timbres. 
      However, different tracks might still have duplicated sequences of notes such as drum patterns typical of a specific music genre.
      Thus, in this section, we evaluate the number of unique drum sequences contained in each dataset, and especially their coverage relative to other datasets, to estimate the possible generalization of the trained model.
      The heatmaps in Figure~\ref{fig:overlap} represent the coverage that a source dataset has on a target dataset in terms of drum sequences. Because the drum sequences are not defined in the datasets, we consider the notes played by the drum instruments in a single beat.\footnote{Specifically, the sequences are identified by computing a "fingerprint" for each beat of the datasets: the notes are mapped down to a common 5-class vocabulary without velocity information; they are quantized into 12 evenly-spaced bins to remove humanization; and collisions between notes are removed (i.e., notes mapped to the same class and quantized to the same position are not counted).} Moreover, because the distribution of beat sequences has a long tail (i.e., many sequences are played a few times while a few sequences are played many times), we are interested in both the coverage in terms of beats and the unique beat sequences between datasets. 
      The reason of this is because while a source dataset might cover only a small portion of the unique beat sequences of any target dataset (Figure~\ref{fig:GroupsOverlap}), those beat sequences may include most of the beats (Figure~\ref{fig:BeatsOverlap}).

      We can observe in the figures that ADTOS has the highest coverage of both beats and unique beat sequences from all the datasets. This is not surprising considering that it has both the highest number of beats (3.2M) and, consequently, the highest number of unique beat sequences (204k). Although a high number of beats does not necessarily indicate diversity, ADTOS has in any case $\approx6\%$ of unique beats, and this is despite the fact that it reuses each drum MIDI loop twice on average. Such a result is higher than in SLAKH and TMIDT, which do not reuse MIDI, but contain offline annotations and have only $\approx3\%$ and $\approx4\%$ of unique beats, respectively.
      On the one hand, the higher diversity of ADTOS largely increases its coverage of unique sequences compared to the other datasets (e.g., ADTOS covers 28\% of the sequences of ADTOF-YT, while TMIDT covers 13\% of them). On the other hand, the high coverage of sequences does not translate to a much higher coverage of beats, because of their unbalanced distribution (e.g., ADTOS covers 94\% of the beats of ADTOF-YT, while TMIDT covers 88\% of them).


\end{appendices}

\bibliography{bibliography}
\end{document}